\documentclass[12pt]{article}
\usepackage{graphicx}
\begin{document}
\begin{center} {\Large \bf Neutrino oscillations in medium with periodic
square potential} \\
\vspace{1cm}
{\large \bf N. A. Kazarian$^1$, M. A. Nalbandyan$^2$} \\
{\it $^1$Institute for Theoretical Physics and Modelling, 375036,
Yerevan, Armenia \\
$^2$Yerevan State University, 375049, Yerevan, Armenia}
\end{center}
\begin{abstract}
We have investigated two flavor neutrino oscillations in medium
with periodic step electron number density profile. An approximate
analytical solution have been found when the length of the density
fluctuation is smaller then the neutrino oscillation length.
\end{abstract}

{\it \large \bf Introduction}

Neutrino oscillation mechanism gives an explanation for decreasing
flux intensity of electronic neutrinos coming from the Sun. The
problem was predicted in \cite{Pontecorvo:1967fh} and was solved
via averaged vacuum oscillations with large mixing angles in
\cite{Gribov:1968kq}. In \cite{Wolfenstein:1977ue},
\cite{Mikheev:1986wj}, \cite{Mikheev:1986gs} solution for neutrino
oscillations in matter with constant electron number density were
obtained. An adiabatic neutrino oscillations were investigated in
\cite{Mikheev:1986if}, \cite{Friedland:2000rn}. An exact analytic
expression  for the probability amplitudes describing the
oscillations of two (flavor) neutrinos in matter with
exponentially varying density was obtained in
\cite{Petcov:1988wv}. This theoretical picture of such potential
is in good agreement with the standard solar model according to
which the matter density in the sun decreases exponentially with
the distance from the center of the sun, except for the regions
located close to the center and to the surface. But such a picture
doesn't take into account the unpredictable variations of
electronic gas density. So it is naturally interesting to studying
the potentials with variate density profile. A potentials with
periodically variate profile were studied independently in
\cite{erm}, \cite{Akhmedov:1988kd} in connection with a
possibility of the parametric resonance in two neutrino system. In
\cite{erm} an approximate solution for sinusoidal matter density
profile was found. In \cite{Akhmedov:1988kd} the exact analytic
solution for the periodic step-function density profile was
obtained. In \cite{Ioannisian:2004jk} an analytic approximate
solution for arbitrary electron density for small matter effects
on the neutrino oscillations was wound.

In the present paper we have constructed toy model of electron
density fluctuations. An approximate solution for two flavor
neutrino oscillations probabilities was found in periodic square potential
profile with suggestion that the length of each square is smaller than the
oscillation length.
\\

{\it \large \bf Neutrino oscillation in the periodic square
potential}\\

In this section, we consider the mixing of two (active) neutrinos.
$\nu _f   = U\nu _{mass} $, where $ \nu _f  = \left(
 \nu _e, \nu _\alpha   \right)^T $ are flavor states, $\nu_\alpha$
 is a linear combination of $\nu_\mu$ and $\nu_\tau$,
 $ \nu _{mass}  = \left( \nu _1, \nu _2 \right)^T $
are mass states with masses \(m_1\) and \(m_2\) respectively.
\begin{equation}
 U
= \left( {\begin{array}{*{20}c}
   {\cos \theta } & {\sin \theta }  \\
   { - \sin \theta } & {\cos \theta }  \\
\end{array}} \right)
\end{equation}
is a lepton mixing matrix in weak charged interactions. The
equation for neutrino oscillations in matter is given by
Schrodinger equation
\begin{equation}
 - i\frac{d}{{dx}}\left( \begin{array}{l}
 \nu _e  \\
 \nu _\alpha   \\
 \end{array} \right) = \left[ {U\left( {\begin{array}{*{20}c}
   { - \frac{{\Delta m^2 }}{{2E}}} & 0  \\
   0 & 0  \\
\end{array}} \right)U^T  + \left( {\begin{array}{*{20}c}
   {V} & 0  \\
   0 & 0  \\
\end{array}} \right)} \right]\left( \begin{array}{l}
 \nu _e  \\
 \nu _\alpha   \\
 \end{array} \right),
\label{shred}
\end{equation}
Here \( V = \sqrt 2 G_F N_e \),  $ N_e $ is electron number
density, \( G_F \) is Fermi constant, \( E \) is the energy of  a
neutrino, \( \Delta m^2  = m_2^2 - m_1^2 \),  \( x \)- the length
of the neutrino beam path.

We will assume that the \(V\) is of the form  periodic
rectangularly shaped potentials, as shown in Fig.1.

\begin{figure}[htbp]
\centerline{\includegraphics[width=3.48in,height=1.97in]{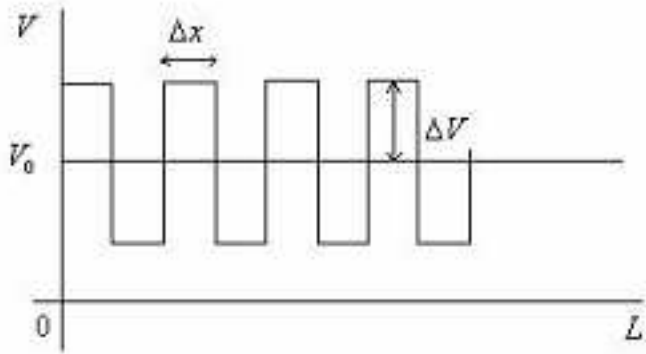}}
\label{fig1} \caption{The dependence of  the electronic potential
\( V\) on distance \( L\). \( V_0\) - is the averaged potential,\(
\Delta V\)is the amplitude of the additional rectangularly shaped
potential and \(\Delta x\) is the half-period of the potential.}
\end{figure}

The formal solution of eq. (\ref{shred}), that is the oscillation
matrix for the flavor states from initial point $x_0$ to the final
point $x_f$, can be written as
\begin{equation}
S_{x_0 \to x_f}= T e^{\int_{x_0}^{x_f}H(x) dx} \ ,
\end{equation}
where $T$ means the chronological ordering.

Thus in order to get the oscillation matrix it is necessary to
multiply consistently solutions on each period. Thus we get the
oscillation matrix in the form of
\begin{equation}
S = \left( {e^{iC} } \right)^n  = e^{inC},
\end{equation}
where $C$ is a hermitian matrix
\begin{equation}
C = \frac{1}{i}ln\left( {e^{iH^ +  \Delta x} e^{iH^ -  \Delta x} }
\right). \label{log}
\end{equation}
Here
\begin{equation}
 H^ \pm   = U\left( {\begin{array}{*{20}c}
   { - \frac{{\Delta m^2 }}{{2E}}} & 0  \\
   0 & 0  \\
\end{array}} \right)U^T  + \left( {\begin{array}{*{20}c}
   {V_0 \pm \Delta V } & 0  \\
   0 & 0  \\
\end{array}} \right)
\ {\rm and}  \ \   \Delta x = \frac{{x_f  - x_0 }}{{2n}}
\end{equation}
is the half period of the potential ,  \(n\) is the number of
periods.

The $S$ matrix can be rewritten in the following form
\begin{equation}
S=e^{inC}=e^{in O^\dagger C^{diag} O} =O^\dagger e^{inC^{diag}} O
=O^\dagger\left( {\begin{array}{*{20}c}
   {e^{  inC^{diag}_{11} } } & 0  \\
   0 & {e^{inC^{diag}_{22} } }  \\
\end{array}} \right)O,
\end{equation}
where $C^{diag}$ is diagonalized $C$ matrix, $C^{diag}_{11}$ and
$C^{diag}_{22}$ are its diagonal elements, $O$ is the matrix that
diagonalize matrix $C$.
\begin{equation}
O = \left( {\begin{array}{*{20}c}
   {e^{ - i\gamma } } & 0  \\
   0 & {e^{i\gamma } }  \\
\end{array}} \right)\left( {\begin{array}{*{20}c}
   {\cos \varphi } & {\sin \varphi }  \\
   { - \sin \varphi } & {\cos \varphi }  \\
\end{array}} \right).
\end{equation}
The phase $\gamma$ is unphysical and does not enter into the
transition probabilities.

Solving these equation we find that
\begin{eqnarray}
\label{sol} \sin ^2 2\varphi  &=&
\frac{{4\left|C_{12}\right|^2}}{{\left( {C_{11}-C_{22}} \right)^2
+ 4\left|C_{12}\right|^2}}, \\
C_{22}^{diag}-C_{11}^{diag}  &=& \sqrt {\left( {C_{11}-C_{22}}
\right)^2 + 4\left|C_{12}\right|^2} .
\end{eqnarray}

Now we can write the probability of $\nu_e-\nu_{\alpha}$
transition:
\begin{equation}
\label{prob}
 P_{\nu _{\rm{e}}  \to \nu _\alpha } = \left| { S _{\nu_e
\nu_\alpha} } \right|^2 = \sin ^2 2\varphi \sin ^2
n\frac{{C_{22}^{diag}-C_{11}^{diag}}}{2}.
\end{equation}

We assume that the period of the potential, $ 2 \Delta x$, is
smaller than the neutrino oscillation length in mater with average
potential $V_0$ and the amplitude of the additional rectangularly
shaped potential $\Delta V $ is smaller than  $(\Delta x)^{-1}$.
In that case we may perform series expansion for  (\ref{log})
(according to the Hausdorf formula)
\begin{equation}
C = C^{(1)} + C^{(2)}  + C^{(3)}  + C^{(4)}  + O(\Delta x^5 ) ,
\end{equation}
where
\begin{eqnarray}
 C^{(1)}  &=& \left( {H^ +   + H^ -  } \right)\Delta x , \nonumber\\
 C^{(2)}  &=& \frac{i}{2}\left[ {H^ +  ,H^ -  } \right]\Delta x^2,\\
 C^{(3)} &=& \frac{1}{{12}}\left[ {\left( {H^ -   - H^ +  }
\right),\left[ {H^ +  ,H^ -  } \right]} \right]\Delta x^3 ,\nonumber \\
 C^{(4)} &=& \frac{i}{{24}}\left[ { {H^ -  } ^2
{H^ +  } ^2  -  {H^ +  } ^2  {H^ -  } ^2 + 2\left( {H^ + H^ - H^ +
H^ -   - H^ -  H^ + H^ - H^ +  } \right)} \right]\Delta x^4
\nonumber.
\end{eqnarray}

The hermitian matrix $C$ has following elements
\begin{eqnarray}
\label{cel}
 &&C_{22}-C_{11} = \frac{\Delta m_m^2 }{E} \cos 2 \theta_m
\Delta x + O(\Delta x^5 )
,\nonumber\\
 &&\rm{  Re} \ {C_{12}}= \frac{{\Delta m_m^2 }}{{2E}} \sin 2
 \theta_m
\Delta x - \frac{{\Delta m_m^2 }}{{12E}} \sin 2 \theta_m \Delta
V^2 \Delta x^3 + O(\Delta x^5 )
,\\
&&\rm{ Im} \ {C_{12}} =  \frac{\Delta m_m^2 }{4E} \sin 2 \theta_m
\Delta V \Delta x^2 + \frac{\Delta m_m^2 }{48E} \sin 2 \theta_m
\Delta V \left[  {\left(\frac{{\Delta m_m^2 }}{{2E}}\right) ^2  -
\Delta V^2 } \right]\Delta x^4 + O(\Delta x^5 ).\nonumber
\end{eqnarray}
Here
\begin{equation}
\label{theta} \sin 2\theta _m  = \frac{{\sin 2\theta }}{{\sqrt
{\left(\cos \theta -\frac{{2EV_0}}{{\Delta m^2 }}\right)^2+\sin^2
\theta } }},
\end{equation}
\begin{equation}
\Delta m_m^2  =   \Delta m^2 \sqrt {\left(\cos \theta
-\frac{{2EV_0}}{{\Delta m^2 }}\right)^2+\sin^2 \theta }
\end{equation}
are the mixing angle and the effective mass square difference in
the medium with the potential $V_0$.

Inserting (\ref{cel}) into (\ref{sol}) and (\ref{prob}) we obtain
an approximate solution for the $\nu_e \to \nu_\alpha$ transition
probability:
\begin{equation}
\label{appsol} P_{\nu _{\rm{e}}  \to \nu _\alpha  } \simeq \sin ^2
2\theta _m \left(1+\beta_1 \right) \sin ^2 \left( {\frac{{\Delta
m_m^2 }}{{4E}} ( x_f  - x_0 ) \left( {1 + \beta_2 } \right)}
\right) ,
\end{equation}
where
\begin{equation}
\label{beta1} \beta_1  = - \frac{1}{{12}} {\Delta V} ^2 \Delta x^2
\cos ^2 2\theta _m + O\left( {\Delta x^4 } \right) ,
\end{equation}
\begin{equation}
\label{beta2} \beta_2  = - \frac{1}{{24}}{\Delta V} ^2 \Delta x^2
\sin^2 2\theta _m + O\left( {\Delta x^4 } \right).
\end{equation}
\\
As it follows from eqs. (\ref{appsol})- (\ref{beta2}) in the
lowest approximation the $\nu_e \to \nu_\alpha$ transition
probability is given by the MSW equation - the oscillations  in
the matter with the average potential $V_0$.

In the first approximation, by the length of the period of the
potential, the oscillation amplitude gets an additional factor
(\ref{beta1}) and the oscillation phase shifts according to
eqs.(\ref{appsol}) and (\ref{beta2}). Thus in that approximation
only ${\Delta V} ^2 \Delta x^2$ enters in the corrections.
According to (\ref{cel}), in the next approximation the
corrections (to the amplitude and to the phase of the oscillation
probability) have forms   $(\Delta V \frac{\Delta m_m^2
}{2E}\Delta x^2)^2$ and $(\Delta V \Delta x)^4$.
\\

{\it \large \bf Conclusion}\\

An approximate analytic solution for two flavor neutrino
oscillations probabilities was found in the periodic square
potential profile with suggestion that the length of each square
is smaller than the oscillation length in the matter with averaged
potential and is smaller than the inverse amplitude ($(\Delta
V)^{-1}$)
of the additional rectangularly shaped potential.\\

{\it \large \bf Acknowledgments}\\

We would like to thank A.N.Ioannisian for suggesting this work to
us, and for many invaluable discussions. We also thank G.Grigorian
for discussions. This work was supported by the National
Foundation of Science and Advanced Technologies under grant No.
ARP2-3234-Ye-04..

\end{document}